\documentclass[twocolumn,showpacs,amsmath,amssymb,pre,aps]{revtex4}   
\usepackage{graphicx}
\usepackage{dcolumn}
\usepackage{bm}

\begin{document}
\title{
Dynamics of Vortex Shells in Mesoscopic Superconducting Corbino Disks
}
\draft

\author{V.R.~Misko and F.M.~Peeters}
\affiliation{
Department of Physics, University of Antwerpen, Groenenborgerlaan 171, B-2020 Antwerpen, 
Belgium
}

\date{\today}

\begin{abstract}

In mesoscopic superconducting disks vortices form shell structures as recently 
observed in Nb disks. 
We study the dynamics of such vortices, driven by an external current $I_{0}$, 
in a Corbino setup. 
At very low $I_{0}$, the system exhibits rigid body rotation while at some critical 
current $I_{c,i}$ vortex shells rotate separately with angular velocities $\omega_{i}$. 
This critical current $I_{c,i}$ has a remarkable non-monotonous dependence on the 
applied magnetic field which is due to a dynamically-induced structural 
transition with a rearrangement of vortices over the shells similar to the 
Coster-Kronig transition in hollow atoms. 
Thermally-activated externally-driven flux motion in a disk with pinning centers
explains experimentally observed $\omega_{i}$ as a function of $I_{0}$ and $T$ and 
the dynamically-induced melting transition. 
\end{abstract}
\maketitle

\section{Introduction}

A mesoscopic Corbino disk is a unique system to study the dynamics 
of self-organized vortex matter in small-size superconductors.
The interplay between the vortex-vortex and vortex-boundary interactions in mesoscopic 
superconductors leads to 
shape-induced giant vortex states \cite{gvs}, concentric {\it shells} of vortices and 
symmetry-induced vortex-antivortex ``molecules'' in mesoscopic squares and triangles 
\cite{vav}.
In the Corbino geometry \cite{crabtree} an applied current creates a gradient in the 
current density 
and thus the Lorentz force, i.e., introducing a shear driving force between the rings 
of vortices. 
This gives us the unique opportunity to study various dynamical effects related to 
vortex motion, e.g., the transition from elastic to plastic motion, 
channeling \cite{chan}, 
vortex friction \cite{fric}, etc. 
The dynamics of self-organized mortex matter in mesoscopic disks has many common
features to, e.g., atomic matter, charged particles in Coulomb crystals, vortices 
in rotating Bose-Einstein condensates, 
magnetic colloids, synthetic nanocrystals, etc. \cite{mitchell,pertsinidis,murray}, 
or even large charged balls diffusing in macroscopic Wigner rings 
\cite{coupier}
and 
can provide us with a deeper understanding of, e.g., the microscopic nature of 
friction, transport, magnetic, optical and mechanical properties of various physical 
and biological systems. 

In a Corbino disk, the 
applied current is injected at the center and removed at the perimeter 
(see Fig.~1) 
to induce a radial current density $J$ that decays as $1/\rho$ along 
the radius \cite{crabtree,paltiel}.
As a result, vortices near the center of the disk experience a stronger Lorentz 
force $F_{L}$ than those near the disk's edge. 
For small $J$, the local shear stress is small and the whole
vortex pattern moves as a rigid body. 
Larger $J$ result in a strong spatially inhomogeneous stress that 
breaks up the vortex solid and concentric annular regions move with
different angular velocities. 
The voltage profiles measured in experiments \cite{crabtree} 
reflect different dynamical phases 
(elastic motion, shear-induced plastic slip) of vortex motion. 
The onset of plasticity in large Corbino disks was theoretically analyzed  
within a continuous model in \cite{marchetti}. 
Within molecular dynamics (MD) simulations of interacting vortices at $T=0$, 
the nucleation and motion of dislocations in the vortex lattice was studied 
in \cite{miguel}. 

Recently, using the Bitter decoration technique, the first direct observation 
of rings of vortices for mesoscopic superconducting Nb disks was reported 
\cite{grigorieva}. 
For 
vorticities $L=0$ to 40, the circular symmetry 
led to the formation of concentric {\it shells} of vortices,
similar to electron shells in atoms or in nano-clusters \cite{kresin}. 
The analysis of shell filling revealed ``magic-number'' configurations (MNC)
\cite{SPB95,BPB94,campbell} corresponding to a commensurability between 
the shells which occurs when the numbers of vortices of each shell have 
a common divider \cite{SPB95}.

Here, we study the dynamics of {\it vortex shells} in {\it mesoscopic} Corbino disks.
Our system has the added flexibility that we have several experimentally accessible 
tuning parameters as, e.g., the driving shear force, the separation between the vortex 
rings (through the external magnetic field) and the commensurability between the 
vortex rings (through the relative number of vortices in each ring). 
In mesoscopic Corbino disks, we 
reveal a non-monotonous dependence of the critical current $I_{c}$ separating 
two dynamical regimes,
a ``rigid body'' or separate shell rotation, 
on the magnetic field $h$, and the appearance of dynamical 
instabilities associated with a jump in $I_{c}(h)$. 
We show that this unusual behaviour is related to a ``structural transition'', 
i.e., an inter-shell vortex transition. 
For {\it non-zero temperature}, 
thermally-activated externally-driven flux motion 
is investigated, 
and 
we
explain the 
observed two-step melting transition in Corbino disks \cite{crabtree}. 

\section{Theory and simulation} 

We place a Corbino disk which has thickness $d$ and radius $R$ in 
a perpendicular external magnetic field $\bm{H}_0$. 
The Corbino setup is shown in Fig. 1. 
An external current flows radially from the center to the edge of the disk
and 
results in the inhomogeneous sheath current density 
$J(\rho) = I_{ext}/2\pi\rho$, 
which makes vortices closer to the center feel a stronger force compared 
to the ones near the edge. 
The Lorentz force (per unit length) acting on vortex $i$, $\Phi_0 \bm{j} \times \hat z$,
resulting from the external current is: 
\begin{equation}
\bm{f}^J_i = \frac{\Phi_0 I_{ext}}{2\pi} \frac{\bm{\rho}_i \times
\hat z }{\rho_i^2} = f_{0} I_{0} \frac{\bm{r}_i \times \hat z}{r_i^2},
\label{eqfj}
\end{equation}
where $\bm{\rho}_i$ is the vortex position, 
$\bm{r}_i = \bm{\rho}_i/R$, 
and $\hat z$ is the unit vector along the magnetic field direction 
which is taken perpendicular to the disk. 
Here
$f_0 = \Phi_0^2 / 2 \pi \mu_0 R \lambda^2 
= 4 \pi\mu_0\xi^2 H_c^2 / R$ is the unit of force, 
$I_{0} = \mu_0 \Lambda I_{ext} / \Phi_0$, 
and $\Lambda = \lambda^2 / d$. 

\begin{figure}[btp]
\begin{center}
\includegraphics*[width=8.0cm]{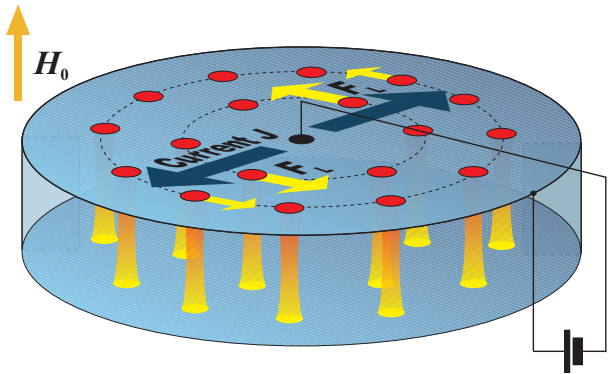}
\end{center}
\vspace{-0.5cm}
\caption{ 
The Corbino setup: the applied current is injected at the center and 
removed at the perimeter of the disk to induce a radial current density 
$J$ (shown by dark blue/dark grey arrows) that decays as $1/\rho$ along 
the radius. 
As a result, vortices near the center of the disk experience a stronger 
Lorentz force $F_{L}$ (shown by yellow/light grey arrows) than those 
near the disk's edge. 
The vortices are shown by red-to-yellow/grey-to-light grey tubes 
and by red/grey spots on the surface. 
The direction of the external applied magnetic field $H_{0}$, which is 
perpendicular to the surface of the disk, is shown by dark yellow/grey 
arrow. 
}
\vspace{-0.5cm}
\end{figure}

In a thin superconducting disk such that $d < \xi \ll R \ll \Lambda$, 
the vortex-vortex interaction force 
$\bm{f}^{vv}_{i}$
and the force of the vortex interaction with the shielding currents and 
with the edge 
$\bm{f}^{s}_{i}$
can be modelled respectively by \cite{buzdin,CBPB04,BCPB04}
\begin{eqnarray}
\bm{f}^{vv}_{i} = f_{0} \sum \limits_{i,k}^{L} 
\left(\frac{\bm{r}_i-\bm{r}_k}{\left|\bm{r}_i
- \bm{r}_k\right|^2}- r_k^2\frac{r_k^2\bm{r}_i - \bm{r}_k}
{\left|r_k^2\bm{r}_i - \bm{r}_k \right|^2}\right), 
\label{eqfvv} \\
\bm{f}_{i}^{s} = f_{0} \left(\frac{1}{1 - r_i^2}- h\right)\bm{r}_i,
\label{eqfs}
\end{eqnarray}
where 
$h = \pi R^2 \mu_0 H_0 / \Phi_0 = (H_0 / 2H_{c2}) (R / \xi)^2$, 
${\bm{r}_i}$
is the position of the $i$th vortex, and $L$ is the number of vortices, 
or the vorticity. 
Our numerical approach is based on the Langevin dynamics algorithm, 
where the time intergartion of the equations of motion is performed 
in the presence of a random thermal force. 
The overdamped equations of motion becomes:
\begin{equation} 
\eta \bm{v}_{i} \ = \ \bm{f}_{i} \ = \ \bm{f}_{i}^{vv} + \bm{f}_{i}^{vp} 
+ \bm{f}_{i}^{T} + \bm{f}_{i}^{d} + \bm{f}_{i}^{s}. 
\label{eqmo} 
\end{equation} 
Here
$\bm{f}_{i}^{d}=\bm{f}_{i}^{J}$
is the driving force (Eq.~(\ref{eqfj})), 
$\bm{f}_{i}^{vp}$
is the force due to vortex-pin interactions \cite{md0157}, 
and
$\bm{f}_{i}^{T}$
is the thermal stochastic force,
obeying the fluctuation dissipation theorem
\begin{equation} 
<\bm{f}_{\alpha, i}(t) \bm{f}_{\beta, j}(t')> =
2\eta\,\delta_{\alpha\,\beta}\,\delta_{i\,j}\,\delta(t - t')k_BT,
\label{sto} 
\end{equation} 
where
Greek and italic indices refer to vector components and vortex labels,
$\eta$ is the viscosity.
The ground state of the system is obtained by
simulating field-cooled experiments \cite{tonomura-vvm}.

\begin{figure}[btp]
\begin{center}
\hspace*{-0.5cm}
\includegraphics*[width=8.0cm]{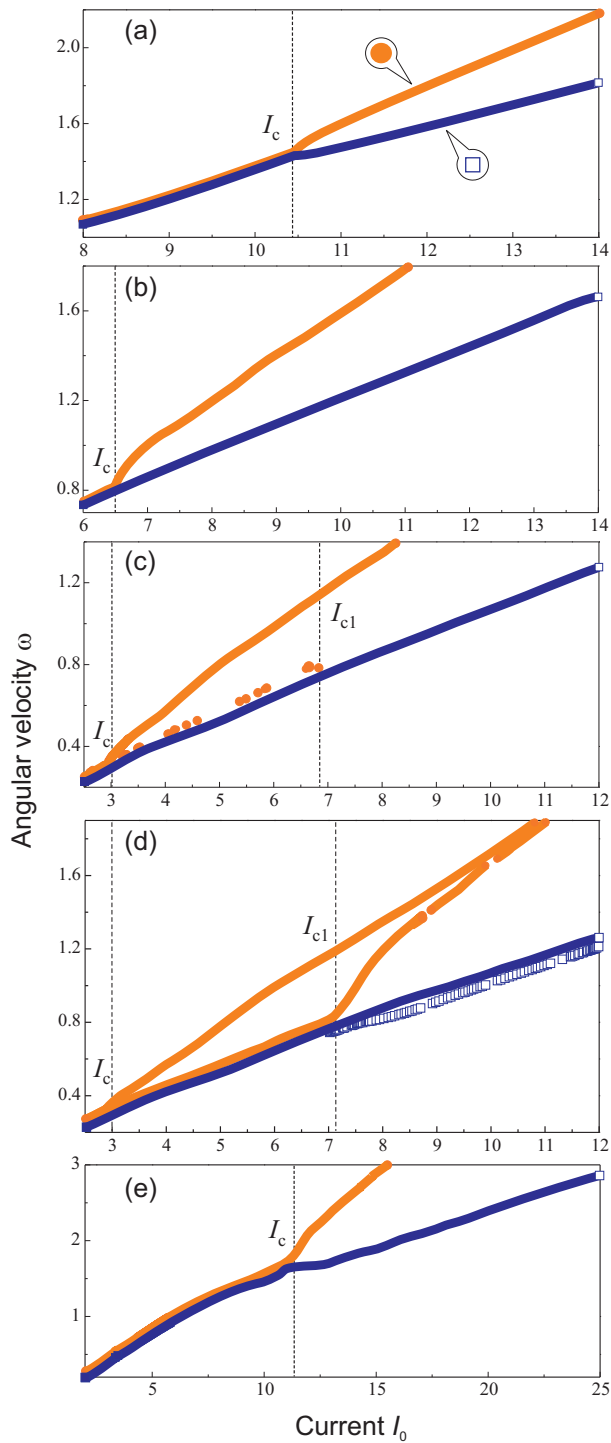}
\end{center}
\vspace{-0.5cm}
\caption{
The angular velocities 
$\omega_{1}$ (orange/light grey solid circles) 
and 
$\omega_{2}$ (blue/dark grey open squares), 
measured in units of $\omega_{0} = 4 \pi \mu_{0} H_{c}^{2} \xi^{2} / \eta R^{2}$, 
of vortices 
in the first and second shells, respectively, 
as a function of applied current $I_{0}$, 
measured in units of $\mu_0 \Lambda I_{ext} / \Phi_0$, 
for 
$L=19$ 
and 
different 
magnetic fields 
$h$: 
(a) $h = 24$,
a rigid-body rotation for $I_{0} < I_{c} \approx 10.4$; 
the shells rotate separately for $I_{0} > I_{c}$;
(b) $h = 25$,
the critical current decreases, $I_{c} \approx 6.5$; 
(c) $h = 26.25$,
the $I_{c}$ further decreases, $I_{c} \approx 3$, 
and the motion becomes unstable; 
(d) $h = 26.5$,
{\it bistable} motion:
a {\it second} critical current $I_{c1} \approx 7.1$ appears; 
(e) $h = 30$, 
the motion stabilizes, and 
first critical current $I_{c}$ disappears, 
and new critical current $I_{c} = I_{c1}$. 
}
\end{figure}

\section{Vortex dynamics in two-shell Corbino disk} 

First, we consider 
the smallest 
mesoscopic 
Corbino disk 
which shows 
the main physics, 
and it has 
$L = 19$
vortices which form the MNC (1,6,12), 
as shown in the left-hand inset of Fig.~3. 
At weak applied magnetic field $h = 24$ all the vortices are packed 
in an almost perfect triangular Abrikosov lattice.
(Here we assume a perfect disk with no pinning, and temperature is set to zero 
after the ``annealing'' process.)
We apply an increasing external current $I_{0}$, and we study 
the average angular velocity $\omega_{i}$ of vortices in each shell.
For small $I_{0}$,
the shear produced by the gradient of the Lorentz force is insufficient to 
break the vortex lattice. 
It produces only {\it elastic} deformations in the lattice which 
rotates as a rigid body with angular velocity 
$$\omega_{\rm RB} = I/2\pi <r^2>,$$ 
where $<r^2> = \sum_{i = 1}^{L} r_i^2 / L$ is the average square
radius for the vortices in the disk. 

With increasing $I_{0}$, the Lorentz force gradient reaches a critical 
value at which the shells start to slide with respect to each other and 
rotate with different velocities 
$\omega_{1} > \omega_{2}$. 
We call this the ``critical current'' $I_{c}$, and for the case 
shown in Fig.~2(a) we found $I_{c} \approx 10.4$. 
Now $\omega_{1} > \omega_{RB} > \omega_{2}$, and 
the inner (outer) shell rotates faster (slower) than the rigid body. 
For higher applied magnetic field, e.g., $h = 25$, 
the critical current decreases, $I_{c} \approx 6.5$ (Fig.~2(b)). 
This is related to a deformation of the shells, which approach a 
circular shape, and therefore can more easily slide. 
At the same time the shells move closer to each other increasing the 
dynamical friction between them. 
At $h = 26.25$, 
the $I_{c}$ further decreases down to $I_{c} \approx 3$, 
and the motion becomes unstable: 
for $I_{0} > 3$, the inner shell rotates with angular velocity 
$\omega_{2} > \omega_{1}$, 
although 
at some values of $I_{0}$ in the region $3 < I_{0} < 6.8$,
$\omega_{2}$ {\it drops} down to $\omega_{2} = \omega_{1}$ (Fig.~2(c)). 
The appearance of this irregular motion, or ``stochastization'', 
is related to the strong dynamical friction between the shells, 
which can even ``lock in'' the shells at some values of $I_{0}$. 
A further increase of the magnetic field, $h = 26.5$ (Fig.~2(d)),
results in {\it bistable} motion of the shells:
a {\it second} critical current $I_{c1} \approx 7.1$ appears, 
and for currents $I_{c} < I_{0} < I_{c1}$ the system 
{\it either} rotates as a rigid body {\it or} the shells 
rotate separately. 
Finally, at even higher field, e.g., $h = 30$, 
the upper branch (corresponding to higher angular velocity) 
of the function $\omega(I_{0})$ for the inner shell 
(shown by orange/light grey dotted line in Fig.~2(d)) 
disappears. 
Therefore, the first critical current $I_{c}$ also disappears, 
and the motion stabilizes with the only critical current 
$I_{c}=I_{c1}$ (Fig.~2(e)). 
The non-monotonous change of the critical current $I_{c}$ with increasing 
applied magnetic field $h$ is summarized in Fig.~3. 
First, $I_{c}$ decreases, then instabilities develop in the system 
which in general are indicative of, and precede, a phase transition. 
The transition to a second critical field occurs through a bistable 
state characterized by {\it two} critical currents in the system. 
This unusual behaviour can be understood by analyzing 
the critical current together with 
the structure of the vortex shells in the region of this ``phase transition''. 
The vortex patterns for magnetic field before (low fields) and after 
(high fields) the transition are presented in the left-hand side and in 
the right-hand side insets of Fig.~3. 
The jump in the critical current is 
correlated to a ``structural transition'' in the system 
where the distribution of vortices over the shells is altered, 
i.e., there is 
an inter-shell vortex transition from a higher orbit to a lower one, 
which is similar 
to the Coster-Kronig transition in hollow atoms (see, e.g., \cite{limburg}).
This structural transition is accompanied by a local re-distribution of the 
flux inside the disk, or by the appearance of flux jump instabilities, which 
might have an additional triggering mechanism caused by the viscosity.

\begin{figure}[btp]
\begin{center}
\vspace*{-0.5cm}
\includegraphics*[width=8.0cm]{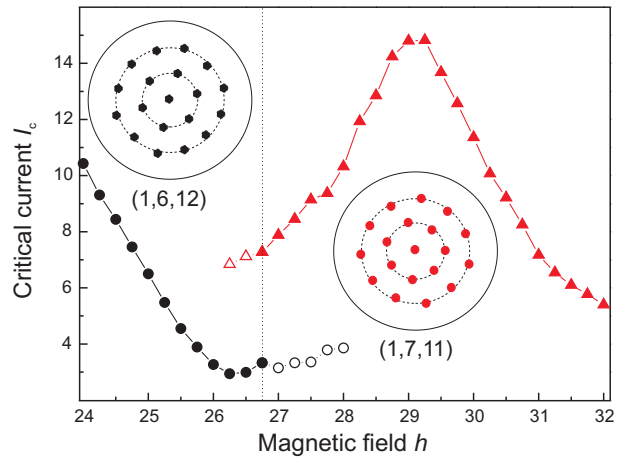}
\end{center}
\vspace{-0.5cm}
\caption{
The critical current $I_{c}$ 
(in units of $\mu_0 \Lambda I_{ext} / \Phi_0$) 
versus 
$h$ 
(in units of $(H_0 / 2H_{c2}) (R / \xi)^2$) 
near the 
structural transition between the states (1,6,12) (solid black circles) 
and (1,7,11) (solid red/grey triangles), insets. 
Empty symbols correspond to bistable vortex motion, 
with two critical currents, $I_{c}$  and $I_{c1}$. 
The jump 
in $I_{c}$
is related to an inter-shell vortex transition 
from a higher to a lower orbit, 
similar to the Coster-Kronig transition in hollow atoms.
}
\vspace{-0.5cm}
\end{figure}

We found that this jump in $I_{c}$ is also observed for other vortex configurations
and is generic for vortex structures confined in Corbino disk. 
For instance, in a disk with $L=22$ it is related to the transition between 
the states (1,8,13) and (2,8,12) at $h=33.5$.
Note that 
a similar behavior (a jump in the mobility) is also found in the 
Frenkel-Kontorova model for the locked-to-sliding transition for chains
moving on commensurate potentials 
\cite{braun}.

\section{Thermally-activated externally-driven dynamics in a Corbino disk} 

Consider now a Corbino disk containing a larger number of vortex shells, 
e.g., $L = 37$ where vortices form three shells.
Note that the chosen vorticity allows the triangular-lattice MNC 
(1,6,12,18) shown in the inset of Fig.~4. 
As in the above case of two shells, 
for small $I_{0}$ the system displays a rigid-body rotation. 
However, the distance between the inner and the outer shells is larger, and, 
thus, the Lorentz force gradient results in a stronger elastic deformation
of the vortex configuration as compared to the two-shells case. 
The shear stress is stronger close to the center of the disk, and as a result, 
the {\it inner shell splits off first} at $I_{c12}$, with increasing $I_{0}$, 
while second and third shells still rotate with the same angular velocity
up to some critical current value $I_{c23} > I_{c12}$ (see Fig.~4).

\begin{figure}[btp]
\begin{center}
\vspace*{0.5cm}
\includegraphics*[width=8.0cm]{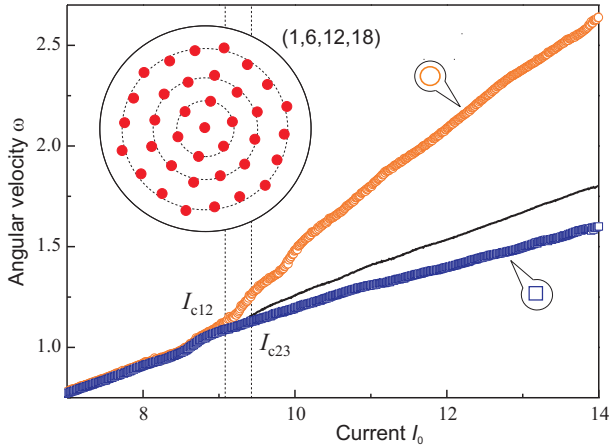}
\end{center}
\vspace{-0.5cm}
\caption{
The angular velocities 
$\omega_{i}$ 
(in units of $\omega_{0} = 4 \pi \mu_{0} H_{c}^{2} \xi^{2} / \eta R^{2}$) 
versus
$I_{0}$ 
(in units of $\mu_0 \Lambda I_{ext} / \Phi_0$) 
in first (orange/light grey open circles), 
second (black solid line) 
and 
third (blue/dark grey open squares) 
shells, 
for a three-shell vortex configuration
with $L=37$.
The inset shows a triangular-lattice ground-state configuration
(1,6,12,18) for $h = 37$ at $I_{0}=0$. 
The critical currents are: 
$I_{c12} \approx 9.1$, and 
$I_{c23} \approx 9.5$.
}
\vspace{-0.5cm}
\end{figure}

The three-shell system displays a number of inter-shell vortex transitions
when changing the applied magnetic field. 
Similar to the case of two shells, 
a structural ``phase transition'' occurs where 
vortices from a higher shell transit to 
a lower shell when the magnetic field increases. 
This leads to non-monotonic dependences of the critical currents 
$I_{c12}$ and $I_{c23}$  on $h$ 
and to instabilities in the functions $\omega_{i}(I_{0})$ 
at different $h$.

\begin{figure}[btp]
\begin{center}
\includegraphics*[width=8.5cm]{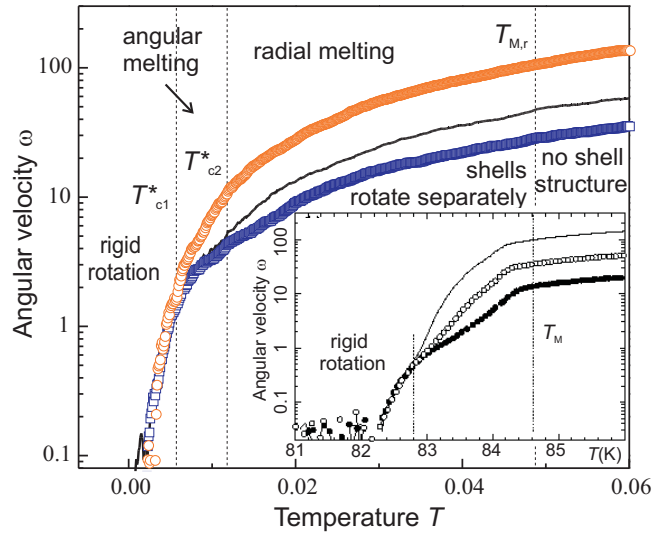}
\end{center}
\vspace{-0.5cm}
\caption{
The angular velocities of vortices 
$\omega_{i}$ 
(in units of $\omega_{0} = 4 \pi \mu_{0} H_{c}^{2} \xi^{2} / \eta R^{2}$) 
versus
$T$ 
(in units of $T_{0} = 4 \pi \mu_{0} H_{c}^{2} \xi^{2} d / k_{B}$) 
in first, 
second 
and
third 
shells (using same symbols as in Fig.~4), 
for 
$L=37$, 
$I_{0} = 0.4$, $f_{p}/f_{0} = 2.5$. 
The onset of 
motion occurs in the form of rigid-body rotation.
At $T = T_{c1}^{\star}$, 
the inner shell splits off, the second shell unlocks at $T = T_{c21}^{\star}$, 
and shells rotate independently ({\it angular} melting). 
The {\it radial} melting, i.e., dissolving of the shells, occurs 
at $T = T_{M,r}$. 
The inset: the experimental $T$-dependence
of $\omega_{i}$ 
obtained from the resistivity measurements at three different distances 
from the center of a Corbino disk \cite{crabtree}.
}
\vspace{-0.5cm}
\end{figure}

Now we will vary the temperature $T$ and 
study the {\it thermally-activated} externally-driven dynamics of vortices 
in the shells.
For this purpose, we introduce pinning centers randomly distributed over the 
disk (for brevety we assume dense narrow pinning centers with maximum pinning 
force $f_{p0}$)
and apply a very low current $I_{0}$ that results in a Lorentz force which is less 
than the pinning force $f_{p0}$ for a vortex in {\it any} shell.
At low $T$ all the vortices are pinned and the shells do not move. 
With increasing $T$, the random-noise force is added to the 
(weak) Lorentz force. 
Intuitively one expects that vortices in the inner shell, 
where the Lorentz force is maximum, 
will depin first and start to rotate while vortices in the other shells are 
still pinned. 
However, the vortex-vortex interaction prevents this scenario, and the 
inner-shell vortices, if unpinned, become ``trapped'' by the vortex lattice 
itself that leads to an elastic deformation. 
The vortex-vortex interaction locks in the vortex configuration, 
and when for further increasing $T$, 
the onset of the motion occurs in the form of a {\it rigid-body} rotation
(Fig.~5). 
At some temperature $T_{c1}^{\star}$, 
the inner shell splits off and starts moving with a higher angular velocity, 
while higher shells stay locked up to a ``second critical'' temperature 
$T_{c2}^{\star}$. 
The transition from a rigid body to a separate rotation of shells is called 
{\it angular melting} (i.e., the ``vortex solid'' to ``vortex shells'' 
transition). This is a multi-step process which starts when first shell 
splits off (i.e., at $T_{c1}^{\star}$ in our three-shell model) and it finishes 
when all the shells rotate independently ($T \geq T_{c2}^{\star}$), but they 
keep their identity and contain a constant number of vortices with well-defined 
average radius $\langle r_{i} \rangle$. 
The {\it radial melting} is associated with a {\it dissolving} of the shells 
(i.e., the ``vortex shells'' to ``vortex liquid'' transition) and occurs at 
a higher temperature, $T_{M,r}$ (Fig.~5). 

This three-shell model explains qualitatively the experimentally observed 
thermally-induced ``solid-liquid'' melting transition 
in a Corbino disk \cite{crabtree} with three probes (see inset to Fig.~4). 
Also, the behaviour of $\omega_{i}(I_{0})$ (Fig.~4) is in agreement 
with the experimental results. 
In the experiment, the superconducting disk contained a large number of 
vortices, and the measurements were done for rings rather than for shells. 
To model larger systems, we used a disk with, e.g., 123 vortices
forming six shells, and we calculated average velocities in three
rings (each containing two shells). 
We found that the 
results for larger disks are similar to the above three-shell system, 
but the radial melting temperature $T_{M,r}$ is lower, 
in agreement with the experiment.

\section{Conclusions} 

We predict an unusual non-monotonous behavior 
of the critical current for unlocking of the vortex rings with magnetic 
field which originates from a ``structural transition'' 
where a vortex jumps from a higher shell to a lower shell 
similarly to the Coster-Kronig transition in hollow atoms. 
Pushing the experiments of Ref.~\onlinecite{crabtree} into the 
mesoscopic regime will allow one to detect experimentally the 
predicted behaviour. 
The vortex motion in the presence of pinning centers reveals the onset 
of a rigid body rotation, 
due to thermally-activated externally-driven flux motion. 
With increasing $T$, the inner shell splits off first and subsequently all the 
shells start moving separately. 
The present numerical study explains the experiments 
of Ref.~\onlinecite{crabtree}, 
in which vortex 
velocities in adjacent layers in a Corbino disk were studied as a function 
of $I$ and $T$. 

\section*{ACKNOWLEDGMENT}

We thank B.~Janko, F.~Nori, and U.~Welp for 
useful discussions and hospitality. 
This work was supported by the Flemish Science Foundation, and the Belgian 
Science Policy. V.R.M. acknowledges a partial support through POD.


\begin{references}

\bibitem{gvs}
V.A.~Schweigert, F.M.~Peeters, and P.~Singha Deo, 
Phys. Rev. Lett. {\bf 81}, 2783 (1998);
A.~Kanda, B.J.~Baelus, F.M.~Peeters, K.~Kadowaki, and Y.~Ootuka, 
Phys. Rev. Lett. {\bf 93}, 257002 (2004).

\bibitem{vav}
L.F.~Chibotaru, A. Ceulemans, V. Bruyndoncx and V. V. Moshchalkov, 
Nature (London) {\bf 408}, 833 (2000); 
Phys. Rev. Lett. {\bf 86}, 1323 (2001); 
V.R.~Misko, V. M. Fomin, J. T. Devreese, V. V. Moshchalkov, 
Phys. Rev. Lett. {\bf 90}, 147003 (2003).

\bibitem{crabtree}
G.W.~Crabtree, D.~Lopez, W.K.~Kwok, H.~Safar, and L.M.~Paulius, 
J. Low Temp. Phys. {\bf 117}, 1313 (1999); 
D.~Lopez, W.K.~Kwok, H.~Safar, R.J.~Olsson, A.M.~Petrean, L.M.~Paulius, 
and G.W.~Crabtree, 
Phys. Rev. Lett. {\bf 82}, 1277 (1999).

\bibitem{chan}
R.~Besseling, P.H.~Kes, T.~Dr\"{o}se, and V.M.~Vinokur, 
New J. Phys. {\bf 7}, 71 (2005).

\bibitem{fric}
A.~Maeda, Y.~Inoue, H.~Kitano, S.~Savel'ev, S.~Okayasu, I.~Tsukada, 
and F.~Nori, 
Phys. Rev. Lett. {\bf 94}, 077001 (2005).

\bibitem{mitchell}
T.B.~Mitchell, J.J.~Bollinger, W.M.~Itano, and D.H.E.~Dubin, 
Phys. Rev. Lett. {\bf 87}, 183001 (2001).

\bibitem{pertsinidis}
A.~Pertsinidis and X.Z.~Ling, Nature {\bf 413}, 47 (2001).

\bibitem{murray}
C.~Murray, C.~Kagan, and M.~Bawendi, Ann. Rev. Mat. Sci. {\bf 30}, 545 (2000).

\bibitem{coupier}
C.~Coupier, M.~Saint Jean, and C.~Guthmann, 
Phys. Rev. E {\bf 73}, 031112 (2006). 

\bibitem{paltiel}
Y.~Paltiel, E.~Zeldov, Y.~Myasoedov, M.L.~Rappaport, G.~Yung, S.~Bhattacharya, 
M.J.~Higgins, Z.L.~Xiao, E.Y.~Andrei, P.L.~Gammel, and D.J~Bishop, 
Phys. Rev. Lett. {\bf 85}, 3712 (2000).

\bibitem{marchetti}
P.~Benetatos and M.C.~Marchetti, Phys. Rev. B {\bf 65}, 134517 (2002).

\bibitem{miguel}
M.-C.~Miguel and S.~Zapperi, Nature Materials {\bf 2}, 477 (2003).

\bibitem{grigorieva}
I.V.~Grigorieva, W.~Escoffier, J.~Richardson, L.Y.~Vinnikov, S.~Dubonos, 
and V.~Oboznov, 
Phys. Rev. Lett. {\bf 96}, 077005 (2006).

\bibitem{kresin}
Y.N.~Ovchinnikov and V.Z.~Kresin, Eur. Phys. J. B {\bf 45}, 5 (2005); 
{\bf 47}, 333 (2005).

\bibitem{SPB95}
V.A.~Schweigert and F.M.~Peeters, Phys. Rev. B {\bf 51}, 770 (1995).

\bibitem{BPB94}
V.M.~Bedanov and F.M.~Peeters, Phys. Rev. B {\bf 49}, 2667 (1994).

\bibitem{campbell}
L.J.~Campbell and R.M.~Ziff, Phys. Rev. B {\bf 20}, 1886 (1979).

\bibitem{buzdin}
A.I.~Buzdin and J.P.~Brison, Phys. Lett. A {\bf 196}, 267 (1994).

\bibitem{CBPB04}
L.R.E.~Cabral, B.J.~Baelus, and F.M.~Peeters, 
Phys. Rev. B {\bf 70}, 144523 (2004).

\bibitem{BCPB04}
B.J.~Baelus, L.R.E.~Cabral, and F.M.~Peeters, 
Phys. Rev. B {\bf 69}, 064506 (2004).

\bibitem{md0157}
C.~Reichhardt, C.J.~Olson, and F.~Nori, 
Phys. Rev. Lett. {\bf 78}, 2648 (1997); 
Phys. Rev. B {\bf 57}, 7937 (1998); 
Phys. Rev. B {\bf 58}, 6534 (1998).

\bibitem{tonomura-vvm}
K.~Harada, O.~Kamimura, H.~Kasai, T.~Matsuda, A.~Tonomura, and V.V.~Moshchalkov, 
Science {\bf 274}, 1167 (1996).

\bibitem{limburg}
J.~Limburg, J.~Das, S.~Schippers, R.~Hoekstra, and R.~Morgenstern, 
Phys. Rev. Lett. {\bf 73}, 786 (1994).

\bibitem{braun}
O.M.~Braun and Yu.M.~Kivshar, {\it The Frenkel-Kontorova model:
concepts, methods and applications} (Springer, Berlin, 2004). 


\end{references}
\end{document}